\documentclass{ieeeaccess}
\usepackage[final]{pdfpages}
\usepackage{epstopdf}
\usepackage{cite}
\usepackage{graphicx}
\usepackage{subfig}
\usepackage{amsmath}
\usepackage{amssymb}
\usepackage{algorithm}
\usepackage{algorithmic}

\usepackage{multirow}
\usepackage[justification=centering]{caption}

\providecommand{\changed}{}

\def\BibTeX{{\rm B\kern-.05em{\sc i\kern-.025em b}\kern-.08em
    T\kern-.1667em\lower.7ex\hbox{E}\kern-.125emX}}
\begin{document}
\history{Received November 18, 2019, accepted November 26, 2019.}
\doi{10.1109/ACCESS.2019.2956698}

\title{Proof-of-Search: Combining Blockchain Consensus Formation with Solving Optimization Problems}
\author{\uppercase{Naoki Shibata}\authorrefmark{1}, \IEEEmembership{Member, IEEE}}

\address[1]{
Graduate School of Information Science, Nara Institute of Science and
Technology, Nara, Japan, 630-0192}

\markboth
{Naoki Shibata: Proof-of-Search: Combining Blockchain Consensus Formation with Solving Optimization Problems}
{Naoki Shibata: Proof-of-Search: Combining Blockchain Consensus Formation with Solving Optimization Problems}

\corresp{Corresponding author: Naoki~Shibata (e-mail: n-sibata@is.naist.jp).}

\begin{abstract}
To address the {\changed large amount of energy wasted} by
blockchains, we propose a decentralized consensus protocol for
blockchains in which the computation can be used to search for good
approximate solutions {\changed to any optimization problem.} Our
protocol allows the wasted energy to be used for finding approximate
solutions {\changed to} problems submitted by any nodes~{\changed (called
  clients)}. Our protocol works in a similar way to {\changed proof-of-work}, and
it makes nodes evaluate a large number of solution candidates to add a
new block to the chain. A client provides a search program that
implements any search algorithm that finds a good solution by
evaluating a large number of solution candidates. The node that
{\changed finds} the best approximate solution is rewarded by the
client. Our analysis shows that the probability of a fork and the
variance {\changed in the} block time with our protocol are lower than
{\changed those in proof-of-work}.
\end{abstract}


%

\titlepgskip=-15pt

\maketitle

\section{Introduction}

\IEEEPARstart{S}{ince} the introduction of
Bitcoin~\cite{nakamoto2008bitcoin}, there have been many
cryptocurrencies built on distributed public ledgers called
blockchains. A blockchain is a growing list of blocks of data items
{\changed that are} designed to be resistant to modification of the
data. In a cryptocurrency, each transaction is registered as an item
in a block. Bitcoin uses a {\changed proof-of-work}~(PoW) system to decide which
{\changed outcome} is the correct outcome of the latest transactions
and prevent double-spending of coins. In a PoW system, the
participating nodes are asked to do some computational task to make a
majority decision on a one-CPU-one-vote basis. While PoW works very
robustly against {\changed misbehavior} and malicious participants on
{\changed a} network where impersonation is easy, the {\changed very
  large} amount of electricity expended {\changed by PoW systems} is
becoming a problem. The estimated energy consumed by Bitcoin was at
least 2.55 gigawatts in 2018, which is comparable to the electricity
consumed in countries such as Ireland~(3.1 gigawatts)~\cite{energy}.

To address this problem, various alternatives to PoW and distributed
public ledgers based on different working principles have been
proposed. Although many of the alternative methods exhibit better
energy efficiency {\changed than PoW}, some of these methods introduce
a single point of {\changed failure,} or they have to trust an
external party. The main advantage of PoW is that it does not need to
trust anything except that it requires a majority of CPU power to be
controlled by honest nodes. There are only a few alternative methods
that satisfy this condition. Bitcoin and cryptocurrencies based on PoW
are still predominant because of {\changed their} unparalleled
security and robustness.

In this paper, we propose a consensus protocol for blockchains, named
{\changed proof-of-search}~(PoS). Our protocol allows computation for making a
consensus to be used for finding a good approximate solution of an
instance of any optimization problem. Our protocol allows a blockchain
to be used as a batch processing system for solving optimization
problems. Any user, called a client, can submit a job for finding a
solution of an optimization problem. Our protocol has functionalities
for submitting a problem instance and delivering the found solution.

The rest of this paper is organized as follows.
Section~\ref{sec:related} introduces related work, including Bitcoin
and {\changed proof-of-work}, along with alternatives to {\changed
  proof-of-work}. Section~\ref{sec:proposed} explains our proposed
consensus protocol, including the main idea, job requests and
execution, environment, and method of compacting a blockchain.
Section~\ref{sec:consideration} discusses properties of our protocol
{\changed including the} requirements for the evaluator, probability
of fork {\changed occurrence}, and variance {\changed in} block
time. This section also describes {\changed the} properties of PoS and
security considerations. Finally, Section~\ref{sec:conclusion}
presents our conclusions, including some thoughts on uses for a
blockchain with our protocol and the benefits {\changed of} using our
protocol.

\section{Related Work}

\label{sec:related}

\subsection{Bitcoin and Proof-of-Work}

\label{sec:bitcoin}

\begin{figure}[b]
\centering \includegraphics[width=0.9\columnwidth]
           {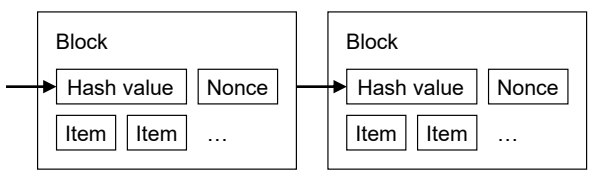}
\caption{Distributed timestamp server in Bitcoin~\cite{nakamoto2008bitcoin}}
\label{fig:timestamp_bitcoin}
\end{figure}

Bitcoin~\cite{nakamoto2008bitcoin} is a robust, secure and
decentralized cryptocurrency. It is built on a peer-to-peer
distributed timestamp server {\changed that generates} computational
proof of the chronological order of transactions. This proof is
provided by a PoW system. The idea of {\changed a} PoW was first
proposed as a way to deter {\changed spam}
e-mails~\cite{dwork1992pricing, jakobsson1999proofs}. In Bitcoin, PoW
is used to enforce majority decision making on a one-CPU-one-vote
basis on peer-to-peer networks where a user can allocate many IP
addresses and one-IP-address-one-vote does not work.

The distributed timestamp server~(Fig. \ref{fig:timestamp_bitcoin}) in
Bitcoin works by forming a linked list of blocks of data items to be
timestamped. {\changed This linked list is called a blockchain.} Each
block contains the hash value of the last block and the data
items. Each time a new block is added to the chain, a hash value of
the new block is computed and widely published. The PoW in Bitcoin
involves finding a value whose hash value begins with a required
number of zero bits. Each block has an entry for an integer value
called a nonce, and only a block that has a nonce that makes the
block's hash value begin with the required number of zero bits is
accepted as a valid block. An incentive is provided to nodes that
support the network {\changed such that} a new coin {\changed is
  given} to a node that succeeds in adding a new block. Network nodes
that try to add blocks are called miners. Honest miners try to add a
new block to the longest chain.  As long as a majority of CPU power is
controlled by honest miners, an honest chain will grow the fastest. In
this way, a majority decision is made. The time required for a block
to be added is called the block time. The required number of zero bits
is automatically adjusted to make the expected block time 10 minutes.

Bitcoin is not a mere online payment {\changed system;} it aims at
{\changed replacing} a currency. For this purpose, the system has to
be extremely {\changed robust, and it} does not depend on any party or
computing node. The Bitcoin entity is the data in the network. It will
never vanish into nothing as long as there are enough honest
nodes. Bitcoin has {\changed the following} properties:

\begin{itemize}
\item Decentralized and self-regulated
\item No need to trust any node or any party
\item Hard to modify the data in the blocks
\item Immune to Sybil attacks
\item The winning probability for each miner is proportional to its
  computational power
\item Legitimacy of blocks can be quickly verified at anytime by any node
\item Any node can join at anytime without {\changed preregistration}
\end{itemize}

However, Bitcoin requires miners to use their computational resources
for PoW, which is basically {\changed the} repeated calculation of
hash values. This {\changed requirement} is a waste of computational
resources and electricity, which could be applied to useful work.

\subsection{Alternatives to Proof-of-Work}

\begin{table*}[tb]
\begin{center}
{
\changed
\footnotesize
\begin{tabular}{c|p{.8\columnwidth}|p{.8\columnwidth}}
\hline
Protocol & Strength & Weakness \\
\hline
Proof-of-work & Extremely robust & Wastes electricity and computational resources\\
Proof-of-stake & No waste of electricity & The richest nodes have control of the network\\
Proof-of-activity & Requires both majority of CPU power and majority of coins to take control of the network & In-between PoW and proof-of-stake\\
Proof-of-burn & No waste of electricity & Nodes that have the right to vote have to be trusted\\
Proof-of-useful-work & Consensus is made by solving {\changed the} Orthogonal Vectors problems & Unclear public demand for solving such problems\\
Primecoin & Consensus is made by searching for chains of prime numbers & Small public demand for finding prime numbers\\
Gridcoin & Rewards miners who perform computation on BOINC & Depends on BOINC\\
Permacoin & Mining resources can be used for distributed storage of archival data & High redundancy of data storage is required\\
Tendermint & No waste of electricity & Utilizes proof-of-stake approach to prevent Sybil attacks\\
Proof-of-space & Allows mining by storing data & Exhibits unique weaknesses\\
Proof-of-luck & No waste of electricity & Requires trusted execution environment\\
Proof-of-elapsed-time & No waste of electricity & Requires trusted execution environment\\
\textbf{Proof-of-search} & Consensus is made by solving optimization problems &
Falls back to PoW if there is no problem to solve\\
\hline
\end{tabular}
\caption{Strengths and weaknesses of consensus protocols}
\label{tab:strength}
}
\end{center}
\end{table*}

To address the wasted computational resources and energy used by
Bitcoin, various alternative consensus protocols for blockchains have
been proposed.

\bigskip

Proof-of-stake and {\changed proof-of-burn}~\cite{slimcoin} are techniques used in
some of {\changed the} recently developed cryptocurrencies.

Proof-of-stake, which {\changed was} first implemented in
Peercoin~\cite{peercoin}, is a consensus protocol that chooses the
creator of the next block based on combinations of random selection
and wealth or age. With this protocol, the node {\changed that} has
more coins will create blocks more often, and thus more coins are
granted. In this protocol, the designated creator of the next block
has to be trusted. The richest nodes are more likely to be selected,
and thus they have control of the network.
Proof-of-activity~\cite{bentov2014proof} is a combination of PoW and
{\changed proof-of-stake}. In this scheme, miners work on PoW to add an empty
block header. In this header, a random group of validators are
designated in the same way as in {\changed proof-of-stake}. These validators are
asked to sign the new block. If the new block is signed by all the
chosen validators, it is added to the chain. The advantage of
{\changed proof-of-activity} is that it requires both a majority of CPU power and
a majority of coins to take control of the cryptocurrency.

Proof-of-burn~\cite{slimcoin} is a consensus protocol that gives a
right to vote to a node if it sends its coin to a special address
where the coin cannot be redeemed. Obviously, nodes that have
{\changed the right} to vote have to be trusted in this scheme. There
is an interesting use of {\changed proof-of-burn that moves} coins
from one cryptocurrency to another cryptocurrency. To do {\changed
  so}, the coins are sent to the address specific to the destination
cryptocurrency where the coins cannot be redeemed with the original
cryptocurrency. Then, a new transaction is added to the destination
cryptocurrency, indicating that the user received the equivalent amount
of coins, {\changed which refers} to the transaction in the original
cryptocurrency.

\bigskip

With {\changed proof-of-useful-work}~\cite{ball2017proofs},
Gridcoin~\cite{gridcoin} and Permacoin~\cite{miller2014permacoin},
computational resources in consensus formation can be used for more
meaningful purposes.

In {\changed proof-of-useful-work}~\cite{ball2017proofs} and
Primecoin~\cite{King:2013:PCP}, the computation for making a consensus
is used for more meaningful purposes than {\changed the} calculation
of hash values. Proof-of-useful-work uses the computation to solve
{\changed the} Orthogonal Vectors problems, while Primecoin makes a
consensus by searching for chains of prime numbers. However, it is not
clear how much public demand there is to find solutions to such
problems.

Gridcoin~\cite{gridcoin} implements a {\changed proof-of-research} scheme, which
rewards miners who perform computations on the Berkeley Open
Infrastructure for Network
Computing~(BOINC)~\cite{Anderson:2004:BSP:1032646.1033223}. {\changed
  The computation for the} consensus protocol is used for scientific
{\changed computations, so} the energy is used for very meaningful
{\changed purposes; however,} the cryptocurrency system has to depend
on the BOINC system. This means that Gridcoin will cease to work if
the BOINC system goes down, and {\changed thus,} Gridcoin is less
robust than Bitcoin.

In Permacoin~\cite{miller2014permacoin}, mining resources are used for
distributed storage of archival data. Successfully minting money in
this system requires random access to a local copy of a file. To mine
coins, a miner needs to prove that the archive file is intact with a
{\changed proof-of-retrievability}, which is an interactive protocol for
cryptographically proving the retrievability of data. {\changed
  However, in order to guarantee access to stored data at any time,
  high redundancy of data storage is required.}

\bigskip

Proof-of-space~\cite{proofofspace} is a protocol between a prover and a
verifier where the prover is supposed to store some large amount of
data. A verifier asks a prover to send a piece of data to check
{\changed whether it is still} storing the data. The protocol is
designed to make the computation, storage requirements and
communication complexity of the verifier small. To use {\changed proof-of-space}
in a decentralized blockchain, a way of determining the winning node
and a way for each miner to know how likely it is to win are
required. The probability of winning should be proportional to the
space allocated for data storage in each node. These practical
considerations are discussed in \cite{Fuchsbauer2015SpacemintAC}. As
mentioned in the paper, {\changed proof-of-space} has its own weaknesses. One of
the problems is that nodes can mine on multiple chains
simultaneously. Miners can also try creating many different blocks
with a single proof-of-space by altering the block contents slightly
and {\changed announcing} the most favorable one.

Proof-of-luck~\cite{milutinovic2016proof} and
{\changed proof-of-elapsed-time}~\cite{sawtoothlake} utilize a trusted execution
environment~(TEE) to form a consensus. A TEE is special hardware that
executes software securely with respect to confidentiality and
integrity. With this kind of hardware, {\changed user interference with} the
consensus process can be avoided, and thus a consensus protocol can be
realized relatively easily. However, only software signed by a trusted
party can be executed on a TEE.

Tendermint~\cite{buchman2016tendermint} is a consensus protocol for
blockchains without mining. This protocol uses a Byzantine fault
tolerance algorithm to form a consensus among a known set of
participants. This is resilient {\changed to} up to $1/3$ of Byzantine
participants being dishonest. This protocol uses a {\changed proof-of-stake}
approach to prevent Sybil attacks.

Nano~\cite{nano} utilizes distributed acyclic graphs~({\changed DAGs})
to store transactions. In a DAG, transactions are stored in nodes
where each node corresponds to a single transaction. A conflict is
resolved by a majority vote among representatives chosen by the
participants. Each vote has a weight calculated as the sum of all
balances of the participants who chose this representative.

\bigskip

The main contribution of this study is that we propose a truly
decentralized consensus protocol for blockchains that allows the
computational power wasted in {\changed proof-of-work} to be used for a more
meaningful purpose than {\changed proof-of-useful-work}~\cite{ball2017proofs} and
Primecoin~\cite{King:2013:PCP}.  Unlike Gridcoin~\cite{gridcoin}, our
protocol does not depend on any party or computing node, and thus, our
protocol is more robust. The working principle of our protocol is
close to that of PoW, and thus it does not need to trust any node to
work correctly.

{\changed The strengths and weaknesses of the consensus protocols are
  summarized in Table \ref{tab:strength}.}

\section{Proposed Protocol}

\label{sec:proposed}

In this paper, we propose a new decentralized consensus protocol for
blockchains, called {\changed proof-of-search}~(PoS), which {\changed has} all the
advantages of {\changed proof-of-work} explained in \ref{sec:bitcoin}. PoS allows
the computational power wasted in {\changed proof-of-work} to be used for
searching for an approximate solution of an instance of {\changed an}
optimization problem. It allows any node to submit a job {\changed
  that includes} a program called an evaluator that evaluates a
solution candidate of a problem to be solved. In our protocol, a
concatenation of a solution candidate and its evaluation value is used
as a nonce instead of an integer. To generate a valid nonce, a node
has to execute the evaluator to evaluate some solution
candidate. Since a large number of nonces have to be generated in the
consensus process, a large number of solution candidates have to be
evaluated. To prevent collusion between nodes, our protocol has two
separate ways of rewarding nodes. A node that succeeds in adding a new
block is rewarded in the same way as in PoW. A node that succeeds in
finding the best solution of an optimization problem is given the
charge paid by the corresponding client. A client can submit a job
without mining, and a miner is not required to submit a job. If no job
is submitted to a PoS-based blockchain, it automatically adds an empty
{\changed job, which} makes PoS work in a similar way to PoW.

In this paper, a solution means an approximate solution. To solve a
problem is to find an approximate solution of the problem.

\subsection{Key idea}

\begin{figure*}[tb]
\centering \includegraphics[width=1.1\columnwidth] {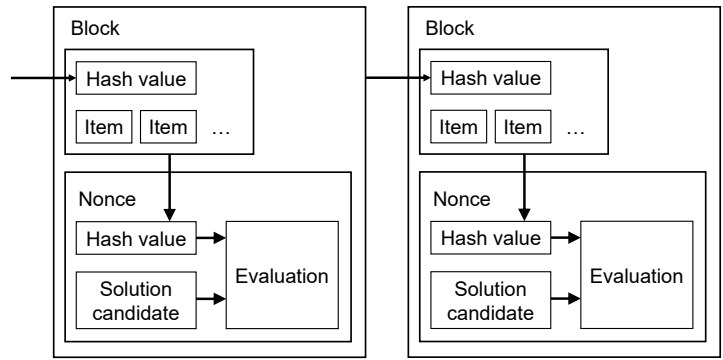}
\caption{Distributed timestamp server in a minimal PoS scheme}
\label{fig:minimal}
\end{figure*}

\label{sec:keyidea}

We first introduce the concepts of an {\it evaluator}, a {\it client}
and a {\it job}. An {\it evaluator} is a computer program that
deterministically computes the evaluation value of a given solution
candidate of an instance of an optimization problem. An evaluator has
to always output the same value if the same input is given regardless
of the platform it runs on. An evaluator includes an instance of a
problem. A {\it job} is data that {\changed represents} an execution
request {\changed for} a search for a solution of an optimization
problem. A job includes an evaluator and all necessary information
regarding the search request. Any node can submit a job to a PoS-based
blockchain, and the node submitting a job is called a {\it client}. A
job can be submitted to the system by registering the job on the
blockchain. The ID of the client is also included in a job. For
example, a client can implement an evaluator to evaluate a solution
candidate of an instance of the traveling salesman problem~(TSP). In
this case, an order for visiting the cities is an input for the
evaluator, and the evaluator outputs the total length of the tour.

In PoS, a concatenation of a solution candidate and its evaluation
value is used as a nonce. A PoS-based blockchain chooses an evaluator
from submitted jobs and specifies which evaluator is used to generate
a valid nonce to add the next block. To add a new block, a mining node
has to find a nonce that makes the block's hash value begin with a
required number of zero bits, in the same way as in {\changed a}
PoW. However, unlike PoW, not every nonce is valid. To generate a
valid nonce, a node executes the specified evaluator to evaluate some
solution candidate. A valid nonce has to contain a solution candidate
and its evaluation value. Since miners have to generate a large number
of hash values to add a new block, we can enforce that miners evaluate
a large number of solution candidates. By requiring a valid nonce that
makes the block's hash value begin with a required number of zero
bits, the system provides a probabilistic proof that the miners have
evaluated a large number of solution candidates. To verify a hash
value means to execute the evaluator with the solution candidate in
the nonce and confirm that the resulting evaluation value matches the
one included in the nonce. Then, the hash value of the block is also
checked to see if it begins with the specified number of zero bits. To
make verification quick, evaluation {\changed also} has to be quick.

There are two objectives for evaluation. One is to find a nonce that
begins with a required number of zero bits. Another objective is to
find a good solution with a good evaluation. In PoS, we assume that a
large number of solution candidates have to be evaluated to find a
good solution. A node that succeeds in finding the best solution among
all nodes will be rewarded by the client. To make this search
efficient, a client provides a computer program called a {\it
  searcher} that implements a randomized search algorithm such as a
genetic algorithm. A searcher is included in a job, and executed by
miners.  Internally, it calls the evaluator many times. Each time an
evaluation is made inside the searcher, it automatically calculates
the hash value of the block to check whether it begins with the
required number of zero bits and {\changed broadcasts} it if it
does. The search continues until a new block is added.

In PoW, there is no need to guard against reuse of the results of
{\changed computations} from the past. This is because the only way to
obtain a hash value is to compute the hash {\changed function and
  because} the ID of {\changed the} miner and the last block are
associated with the hash value. In PoS, however, we need to guarantee
that an evaluation is made each time a block's hash value is
generated. Because the amount of computation in an evaluation can be
substantially larger than that for calculating a hash value, miners
might try to reuse the results of evaluations by sharing them among
different nodes. To prevent this, we have to associate the result of
evaluation with the ID of the miner and other data. To do this, we
make the evaluator take the hash value of all the items in a block
except the nonce itself as the second argument. A tiny amount of error
is introduced in the output of an evaluator to make the output depend
on the second argument. The algorithm for introducing this error has
to be devised and implemented differently by each client. This is like
using an evaluator as a substitute for a hash {\changed
  function. However,} such a property is not strictly required for an
evaluator. If an evaluator is executed twice with the same solution
candidate given for the first argument and different values for the
second argument, then it should be infrequent that the same value is
{\changed returned. However,} this is not strictly prohibited. This
will be further discussed in \ref{sec:loosely}.

Fig.~\ref{fig:minimal} shows the data structure of a blockchain with a
minimal PoS scheme. This minimal scheme works with a single fixed
evaluator, and it does not have a functionality {\changed for
  searching} for a good solution. A blockchain with this scheme works
in a similar way to a blockchain with PoW. In the next subsections, we
describe how to enhance this minimal scheme to add functionalities for
submission of jobs, payment and execution of multiple jobs.

\subsection{Placing a job request}

\label{sec:request}

We want the following three properties in PoS.

\begin{itemize}
\item A miner has a financial incentive to find and provide a good
  solution.
\item A node is not incentivized to submit a problem instance for
  which it already knows a good solution.
\item Submitting a problem instance that is not worth solving is
  financially discouraged.
\end{itemize}

{\changed We especially} need to prepare for the case where a client
knows a good solution for its job. {\changed The possible motivations}
for submitting such a job {\changed are} listed below.

\begin{enumerate}
\item It is advantageous in adding a block.
\item It is rewarded by minted coin.
\item The node can make a profit by finding a good solution.
\end{enumerate}

We need to give honest miners a financial incentive to find a good
solution while preventing malicious miners from making a profit. In
PoS, this is realized by making a client pay {\changed a} charge for
its job. Finding the solution with the best evaluation is only
rewarded by this charge, and in this way, PoS {\changed will have} all
the properties listed above.  For item 1, knowing or finding a good
solution is not advantageous in adding a block, as explained in
\ref{sec:keyidea}. For item 2, finding a good solution is not rewarded
by minted coins. For item 3, a client cannot make a profit by finding
a good solution with the job it submitted. Since a client has to pay
the charge for its job, {\changed a client will only submit} a problem
instance that is worth the charge.

We want to make the payment process fully automated without trusting
any node. To ensure {\changed that the charge is} paid, it is
collected before {\changed the} execution of a job. A client first
submits a job including the charge. Then, the PoS-based blockchain
automatically deducts the charge before execution of the job and pays
it to the winner after completion of the job.

The found solution has to be sent to the client. If a node simply
broadcasts its solution to the network, this solution can be stolen by
another node. If a node encrypts the solution with the public key of
the client, the corresponding private key has to be published
afterward, and the client has to be trusted to do that. To
make sure that the node that found the best solution is paid
automatically without having the solution stolen by other nodes {\changed and without}
trusting any node, each node first registers the evaluation value of
the found solution and the hash value of a concatenation of its ID and
the solution on the blockchain. Then, in the next block time, each
node checks {\changed whether} its solution is the best. The winning node registers
its solution on the blockchain, and then the charge is paid to the
winning node after confirming that the solution is genuine.

\subsection{Simultaneous execution of multiple jobs}

\begin{figure*}[tb]
\centering \includegraphics[width=0.7\textwidth]{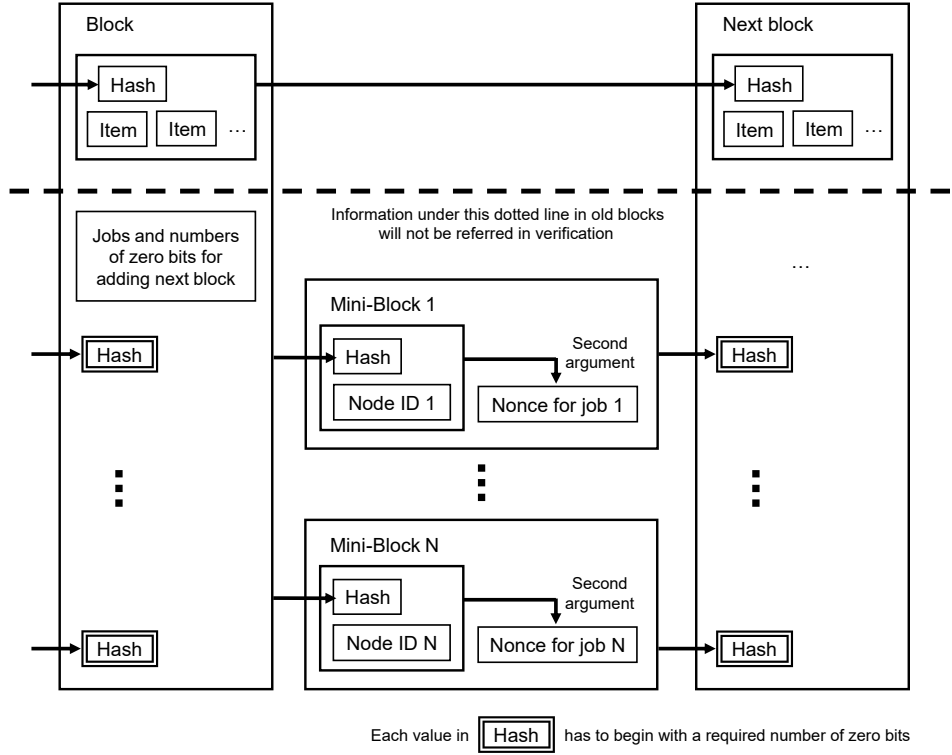}
\caption{Distributed timestamp server in the proposed method}
\label{fig:timestamp}
\end{figure*}

\label{sec:simultaneous}

To make the charge reasonable, we make miners execute multiple jobs at
a time. We also want to make the winning probability of each node
proportional to the computational power spent for the job. A simple
way of realizing these properties might be adjusting the block time
according to the job size. However, this method has {\changed the
  problem of an} increased probability of {\changed fork occurrence}
because the probability depends on the block time. In order not to
increase the probability of fork {\changed occurrence}, we add
{\changed miniblocks} between blocks without changing the block
time. {\changed Each miniblock} corresponds to a job.

A {\changed miniblock} consists of only a nonce, the ID of {\changed
  the} mining node, and the hash value of the last block, as shown in
Fig.~\ref{fig:timestamp}. Miners try to find a valid nonce of any
{\changed miniblock} that makes the {\changed miniblock}'s hash value
begin with the required number of zero bits. As explained in
\ref{sec:keyidea}, a valid nonce contains a solution candidate
{\changed for} the corresponding job and its evaluation value. Each
time such a nonce is found, the node adds the corresponding {\changed
  miniblock} by broadcasting the {\changed miniblock} with that
nonce. A block is added when all the {\changed miniblocks} are added
for all the jobs specified by the PoS-based blockchain. New coins are
{\changed awarded} to all the nodes that add the {\changed
  miniblocks}. Verification of hash values {\changed requires repeating}
the process explained in \ref{sec:keyidea} for each {\changed miniblock}.

Because of message delivery delay in a network, there could be a
difference in the sets of messages received by two different
nodes. Consequently, two mining nodes may have different sets of items
to be included in a new block. Thus, two different nodes may be
executing the jobs to add different blocks. In other words, these
nodes are executing their jobs to add {\changed miniblocks} that contain the
hash values of different blocks. To prevent a mini-fork and
make a larger number of nodes work to add the same block, we make
miners execute jobs on the longest chain whenever possible. Here, one
chain is longer than another if it has more blocks. If two chains have
the same number of blocks, the chain with more {\changed miniblocks} after the
last block is longer. When a node receives a new {\changed miniblock} added by
another node, it checks {\changed whether} the chain associated with the new
{\changed miniblock} is longer. If this chain is valid and longer, the node
immediately starts executing jobs on it. When a node adds a new
{\changed miniblock}, it also broadcasts the corresponding block and all the
added {\changed miniblocks} that come after the block.


We make the expected amount of computation for each job proportional
to the charge. It is fairly straightforward to realize this by making
the PoS-based blockchain adjust the required numbers of zero bits in
the hash values according to the charge. We now assume that the same
amount of computation is required in an evaluation for each job. The
system knows the average total charge $C$ of incoming job requests per
block time by scanning the requests placed in the past. The system
also knows the average number $E$ of evaluations made by all miners
per block time by checking the block times and the numbers of zero
bits of the past {\changed miniblocks}. Let $c_j$ denote the charge for job
$j$. The system should choose the combination of jobs to satisfy
$\sum_{j} c_j \approx C$. Then, the number $z_j$ of zero bits for job
$j$ should be set to satisfy $2^{z_j} \approx c_j \cdot E / C$.
{\changed
The node that succeeds in finding the nonce that makes a miniblock's
hash value begin with the required number of zero bits is rewarded in
the same way as in PoW. However, the total amount of reward for adding new
miniblocks in a block time has to be kept constant, which we denote
by $R$. In order to do this, the reward for adding the miniblock
corresponding to job $j$ is set to $R \cdot 2^{z_j} / \sum_{k}
2^{z_k}$.  }




\subsection{Execution environment}

\label{sec:execution}

To implement PoS, an execution environment is needed for evaluators
and searchers. To make the amount of computation for each job
proportional to the charge, we need to compensate for the difference
in the amount of computation of evaluation for each job. For example,
if the evaluator for job $j_1$ takes two times {\changed the execution
  time that} the evaluator for job $j_2$ {\changed takes,} then 1 more
zero {\changed bit} should be required in the hash value with
$j_2$. Since execution of {\changed the} evaluator has to be
deterministic, we need a platform-independent way of counting the
number of steps. The number of steps can be, for example, the number
of bytecode instructions executed on a virtual machine. The execution
environment has to have a functionality to count the number of steps
in {\changed the} execution of an evaluator. The required number of
zero bits for each {\changed miniblock} has to be adjusted according
to the measured number of steps in {\changed the} execution.

Since any user can submit a job, an evaluator can be inappropriately
implemented. In case {\changed the} evaluator takes too many steps for execution,
there must be a way to terminate this execution after a specified
number of steps. To make execution of {\changed the} evaluator
deterministic, this step count has to be exact. If an evaluator
crashes or is terminated, it is regarded as returning the worst
evaluation.

It is possible that a searcher takes too much computation
compared to the evaluator. This can be easily detected by checking the
number of steps. In this case, a miner is allowed to switch the search
algorithm to a simple random search. By doing this, the node can
increase its hash rate while it would be less likely to find the best
solution.

In summary, the execution environment has to satisfy the following
conditions.

\begin{itemize}
\item It allows safe execution of untrusted code.
\item It provides a way to guarantee that an evaluator runs
  deterministically.
\item It counts the number of steps of execution in a
  platform-independent way.
\item It returns the number of steps after execution.
\item It terminates execution after a specified number of steps.
\end{itemize}

Implementing an interpreter-based virtual machine for the execution
environment satisfying all the above conditions should be
straightforward. Execution can be made deterministic by not providing
APIs that make execution non-deterministic.





\subsection{Compacting blockchain}

\label{sec:compacting}

With the method explained above, all the evaluators recorded in a
blockchain have to be executed to verify a chain. However,
the size of an evaluator {\changed will} be significantly larger than a hash
value, and the storage space for keeping all the evaluators can become
a heavy burden in managing a blockchain. The required storage size can
be reduced by relaxing the requirements {\changed for} verification of a chain. If
the participants think {\changed that} it is sufficient to verify a certain number of
blocks, then the information regarding the jobs recorded in the
older blocks can be discarded. In Fig.~\ref{fig:timestamp}, only the
hash value above the dotted line is checked during verification of old
blocks, and therefore the information below the dotted line can be
discarded for old blocks. Even with this relaxed {\changed method} of verification,
it is very hard to modify the items in the old blocks without redoing
all the PoS for the new blocks.

\begin{algorithm}[b]
\caption{Place a job execution request}
\label{algorithm:request}
{\fontsize{9}{10}\selectfont
\begin{algorithmic}[1]
\REQUIRE Job request $q$
\ENSURE Best found solution $s$
\item[]
\STATE Register $q$ to the blockchain.
\STATE Wait until a solution $s$ is received.
\RETURN{$s$}
\end{algorithmic}
}
\end{algorithm}

\begin{algorithm}[tb]
\caption{Mine}
\label{algorithm:mine}
{\fontsize{9}{10}\selectfont
\begin{algorithmic}[1]                

\STATE Wait until a new block is added. \label{line:initBegin}

\STATE Set $activeChain$ to the added
block. \label{line:initEnd} \label{line:initActiveChain}

\WHILE{true} \label{line:outerLoopBegin}

\STATE \COMMENT{Process for the jobs that will be executed in the next
  block time} \label{line:bt1Begin}


\STATE Create the list of valid unexecuted jobs. \label{line:unexecuted}

\STATE Choose the jobs that will be executed in the next block
time.

\STATE Deduct the charge of these jobs from the client.

\STATE Include these jobs in the block being added. \label{line:bt1End} \label{line:include}

\item[]

\STATE \COMMENT{Process for the jobs that is executed in the current
  block time}

\STATE Verify newly received items~(transactions) and put them in the
block being added. \label{line:newitem} \label{line:bt2End}

\item[]

\STATE \COMMENT{Process for the jobs that were executed two block
  times before}

\STATE Check the blockchain to see {\changed whether} the solutions found two block
times before were the best solutions, as explained in
\ref{sec:request}. Register the solutions to the blockchain if they
were the best.\label{line:bt3End}

\item[]

\STATE \COMMENT{Process for the jobs that were executed three block
  times before}

\STATE Check {\changed whether} the registered solutions are genuine. Process payment
of the corresponding charge.\label{line:bt4End}

\item[]

\REPEAT \label{line:innerLoopBegin}

\STATE Execute one of unfinished jobs listed in the previous block,
until a new {\changed miniblock} is received. \label{line:execute}

\STATE Register the hash value of {\changed the} found solution to the blockchain~(see
\ref{sec:request}). This hash value will be included in the next
block.

\STATE Verify {\changed that} the chain associated with the received {\changed miniblock} by
calling Algorithm \ref{algorithm:verify}. \label{line:verify}

\IF{the {\changed miniblock} is valid and the corresponding chain is
  longer} \label{line:checkChainLength}

\STATE Change $activeChain$ to the chain with the {\changed miniblock}.

\STATE Move the items in orphaned blocks to the list of newly received
items. \label{line:orphan}

\ENDIF

\UNTIL{a new block is added} \label{line:innerLoopEnd}

\ENDWHILE \label{line:outerLoopEnd}


\end{algorithmic}
}
\end{algorithm}

\begin{algorithm}[tb]
\caption{Verify}
\label{algorithm:verify}
{\fontsize{9}{10}\selectfont
\begin{algorithmic}[1]
\REQUIRE {\changed Miniblock} $b$
\ENSURE {\changed Verifies} the chain corresponding to $b$ and returns $true$ iff it succeeds.
\item[]
\FOR{all blocks $k$ in the chain pointed by $b$ in chronological order}
\IF{the same block as $k$ is included in a previously verified chain}
\STATE {\bf continue}
\ENDIF
\STATE Check all the transactions in $k$ {\changed and} return $false$ if it fails.
\IF{$k$ is old}
\STATE Do relaxed verification explained in \ref{sec:compacting} and
return $false$ if it fails.
\ELSE
\FOR{all {\changed miniblocks} $m$ between $k$ and the next block}
\IF{the same {\changed miniblock} as $m$ is included in a previously verified chain}
\STATE {\bf continue}
\ENDIF
\STATE Execute the evaluator and check {\changed whether} the nonce and all the hash
values satisfy the corresponding requirements {\changed and} return $false$ if it
fails.
\ENDFOR
\ENDIF
\ENDFOR
\RETURN{$true$}
\end{algorithmic}
}
\end{algorithm}

\begin{algorithm}[tb]
\caption{Random search}
\label{algorithm:randomsearch}
{\fontsize{9}{10}\selectfont
\begin{algorithmic}[1]
\REQUIRE Hash $h$ of the block to be added, node ID $id$, evaluator $ev$
\ENSURE The best solution candidate found in the search
\item[]
\STATE $bests := null, beste = null$
\REPEAT
\STATE Generate a random solution candidate $s$.
\STATE $e := ev(s, hash(h, id))$
\IF{$e$ is better than $beste$}
\STATE $beste := e, bests := s$
\ENDIF
\STATE $h := hash(h, id, s, e)$
\IF{$h$ begins with the required number of zero bits}
\STATE broadcast the {\changed miniblock}.
\STATE {\bf break}
\ENDIF
\UNTIL{a new {\changed miniblock} is received}
\RETURN{$[bests, beste]$}
\end{algorithmic}
}
\end{algorithm}

\subsection{Putting them all together}

Our protocol is an enhancement of PoW, and {\changed thus, it uses common techniques}
with PoW. A blockchain with our protocol is structured as a
peer-to-peer network. The entire network is loosely connected without
a fixed topology. In order for a node to join a network, it has to
know one of the nodes that is already part of the network. Each node
connects to several random nodes. A message is broadcast with a gossip
protocol. Each node retains a copy of the entire information of a
blockchain.

Our protocol allows any client to submit a job and receive the
solution. This is very simple from the point of view of a client, as shown in
Algorithm \ref{algorithm:request}.

{\changed Executing a job and making} the resulting payment {\changed
  take} at least 4 block times. The following is how a job is
processed in the fastest scenario.

\noindent
\textbf{Block time 0} A job is broadcast by the client.

\noindent
\textbf{Block time 1} Validity of the job is inspected. The charge is
deducted from the client's account. This job is chosen for execution
in block time 2.

\noindent
\textbf{Block time 2} The job is executed.

\noindent
\textbf{Block time 3} Each node broadcasts the hash value of {\changed
  the} found solution. These hash values are registered on the
blockchain.

\noindent
\textbf{Block time 4} The charge is paid to the node that found the
best solution.

Job execution and payments are all processed by miners. Miners process
jobs in a pipelined manner. Algorithm \ref{algorithm:mine} shows how
they are processed from the miner's point of view.

There are two threads running in parallel, and Algorithm
\ref{algorithm:mine} runs on one of them. In the other thread, the
received items are enqueued in the list of newly received items.  To
register an item on the blockchain, it has to be broadcast to the
network. When these items are received by a miner, they are enqueued
in the list.

After Algorithm \ref{algorithm:mine} starts, the miner first {\changed chooses}
the chain to work on~(line
\ref{line:initBegin}-\ref{line:initEnd}). There can be multiple valid
chains, and they all begin with the same block. One specific sequence
of block, or a chain, can be specified by the last block. The miner
starts working on the chain associated with the first block received.

At line \ref{line:unexecuted}, the balance of the client's account is
checked.

At line \ref{line:include}, to make all the miners work on
the same set of jobs, a consensus has to be made on the set of jobs
and the corresponding numbers of zero bits before the miners start
working on them. 

At line \ref{line:execute}, a searcher is executed in the execution
environment explained in \ref{sec:execution}. As an example of a
searcher, a random search algorithm is shown in
Algorithm~\ref{algorithm:randomsearch}. As explained in
\ref{sec:keyidea}, this searcher internally calls the corresponding
evaluator. Each time an evaluator is called, it automatically creates
the nonce with the result of evaluation and checks the hash value {\changed to see} if
it begins with the required number of zero bits. If it does, the new
{\changed miniblock} is broadcast, and the execution of {\changed the} searcher is
terminated. The execution of {\changed the} searcher also terminates if a new
{\changed miniblock} is added by another mining node. Please note that when a
new block is added, the last {\changed miniblock} for that block is also added.

At line \ref{line:verify}, the chain associated with a {\changed miniblock} is
verified. This procedure is shown in Algorithm
\ref{algorithm:verify}. This algorithm works as explained in
\ref{sec:simultaneous} and \ref{sec:compacting}. There is no need to
verify blocks and {\changed miniblocks} if exactly the same blocks or
{\changed miniblocks} are included in a chain previously verified by that node.

At line \ref{line:orphan}, the items contained in orphaned blocks are
moved to the list of newly received items. An orphaned block is a
block that was a part of the chain worked on by the miner, but
{\changed is no longer a part of the chain} because the miner is
{\changed now} working on another chain.



\subsection{Illustrative example}

We now explain how a blockchain with PoS can be utilized from
{\changed the users' point} of view.

Suppose that a client {\changed has} an {\changed instance of an} optimization
problem to be solved. He implements an evaluator and a searcher for
the problem and decides the amount of charge to pay for solving this
problem. He then {\changed constructs} a job by combining the evaluator, the
searcher and the charge and submits this job as explained in
Algorithm \ref{algorithm:request}. The charge is automatically
withdrawn after registering the job. The expected amount of
computation for this job is proportional to the charge. Miners {\changed work}
for the blockchain to find a good solution to the problem. The found
solution is eventually registered to the blockchain, and thus, the
client will obtain the found solution in exchange for the charge.

We assume that there are always many miners trying to add a new block
to the chain. Suppose that a new block time for the longest chain has
just started. The chain provides multiple jobs to work on, and a miner
chooses a job from {\changed these}. If there is no job submitted by
clients, it falls back to PoW, and empty jobs for finding a nonce that
makes the block's hash value begin with the required number of zero
bits are automatically inserted. The miner extracts a searcher and an
evaluator from the job and executes the searcher. Upon execution of
the searcher, the searcher calls its evaluator many times, and the
searcher generates and evaluates many solution candidates. When a
miner finds a good solution candidate, it keeps that solution
candidate for later use. Each time a miner evaluates a solution
candidate, it also generates a valid nonce, and the miner calculates
the hash value of the {\changed {\changed miniblock}} including this
nonce. If the hash value begins with the required number of zero bits,
then the miner broadcasts the {\changed miniblock} containing the
found nonce, in addition to the last block. A miner that succeeds in
adding a new {\changed miniblock} is rewarded in the same way as in
PoW. When a miner notices that a new {\changed miniblock} is added,
the miner starts working on a new {\changed miniblock} that comes
after the previous block broadcast with the new {\changed miniblock}
in order to prevent a mini-fork. A block time continues until all the
{\changed miniblocks} are added. After a block time ends, a miner
registers his best found solution to the blockchain. The
node {\changed that} succeeded in finding the best solution for a job
is given the charge included in the job.

Now, suppose that a client and a miner are colluding to obtain
{\changed an} unfair amount of coins. In order for this miner to add a
new block, {\changed he} has to find a valid nonce that makes the hash
value begin with the required number of zero bits. This is essentially
picking a random solution candidate and calculating the hash
value. The problem and its solution do not matter for this, and
therefore there is no merit {\changed in} a client and a miner to
{\changed colluding}. On the other hand, a client can carry out
computation for its job before submitting it and tell the solution to
a {\changed colluding} miner. However, there is no {\changed benefit}
for doing this since the resulting reward is paid by the client to the
{\changed colluding} miner. This is effectively the client sending its
coin {\changed on the colluding} miner, which is a normal transaction
{\changed in} a cryptocurrency.

{\changed
\bigskip
We now summarize the flow of coins and the incentive mechanism.

\textbf{Incentive for finding the best solution to a job :} Each
client pays the charge when it registers its job. This charge is
automatically withdrawn upon registration and kept by the system
until the completion of the job. The registered job is eventually
executed by miners. After execution of the job, the system knows which
miner found the best solution for each job. The system pays the kept
charge to the account of the miner who found the best solution.

\textbf{Incentive for adding a new miniblock :} The system decides
the required number of zero bits in the hash value for each {\changed miniblock}
according to the charge paid for the job, as explained in
\ref{sec:simultaneous}. When a block time begins, each miner chooses a
{\changed miniblock} to work on, and executes the corresponding searcher. A
large number of nonces are generated upon execution of the searcher.
Each time a nonce is generated, the miner checks whether the hash value of
the {\changed miniblock} including the generated nonce begins with the required
number of zero bits. If it does, the miner succeeds in adding a new
{\changed miniblock}, and the miner receives the reward for the new
{\changed miniblock}. The amount of this reward is set according to the required
number of zero bits in the hash value, as explained in
\ref{sec:simultaneous}.  }

\section{Consideration}

\label{sec:consideration}

\subsection{Requirements for evaluator}

\label{sec:loosely}

The whole scheme of PoS depends on the way it enforces an evaluation
each time a hash value is generated.

We now consider an attack where one of the items in
Fig.~\ref{fig:minimal} is altered by a miner. Altering an item in a
block is not always easy, but here we assume a pessimistic scenario in
which the miner can alter an item freely. Suppose that an evaluator
returns the same value at a rate of once per $u$ times if the same
first argument and a different second argument are given. The miner
could try calculating the hash value of a block faster than it should
in the following way.
\begin{enumerate}
\item The miner alters the item.
\item Calculate the hash value of the block assuming that the
  evaluator returns the same value.
\item Check {\changed whether} the hash value begins with the required number of zero
  bits.
\item If it does, then execute the evaluator to check {\changed whether} it returns
  the assumed value.
\end{enumerate}
The above attempt succeeds at a rate of once per $u$ times. The rate
of generating the valid hash value of a block is higher with this
method if evaluation is very slow and $u$ is small. To
prevent it from paying off, the following condition has to be
satisfied.

\[
u > \frac{\text{\small Amount of computation in evaluation and hash
    calculation}}{\text{\small Amount of computation in hash calculation}}
\]
\bigskip

\subsection{Probability of fork \changed occurrence}

\label{sec:fork}

\begin{figure}[b]
\centering \includegraphics[width=0.8\columnwidth] {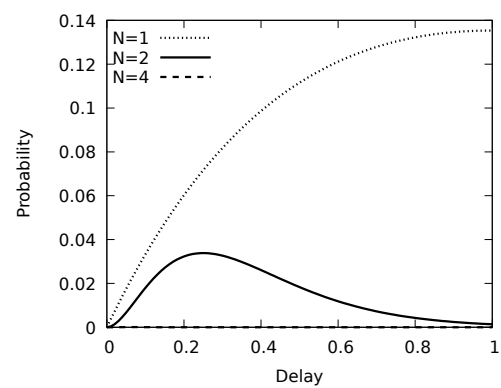}
\caption{Probability of a fork after a block is added, with $N$ {\changed miniblocks}}
\label{fig:forkgraph}
\end{figure}

A fork in {\changed a} blockchain is a situation where there are two
or more valid chains with the same length. In Bitcoin, a fork
{\changed occurs} when a mining node adds a block before knowing that
another node has already added a block. This can happen because of
message delivery delay in a network.

Now, we discuss how the probability of fork {\changed occurrence} is
affected by introducing {\changed miniblocks}. For the sake of
simplicity, we suppose that there are only two mining nodes in the
network. A communication channel with a constant delay $d$ connects
them. We can model block creation as a Poisson process. Now, suppose
that node $A$ has {\changed just} created a block right now. Then, the
probability $p_1$ of a fork {\changed occurring} within time $d$ is
the probability that the other node $B$ creates another block within
this time period, while $A$ does not create a block in this period.

\[
p_{1} = e^{-\lambda d} (e^{-\lambda d} \lambda d),
\]
\noindent
where $\lambda$ is the average number of block {\changed creations}
per interval $\cdot$ node.

We now assume that all {\changed miniblocks} require the same expected amount of
computation to be added. The probability $p_N$ of a fork with length
$N$ after a creation of some block is the probability {\changed that this event
happens} $N$ times in succession.

\begin{equation}
p_{N} = \left \{e^{-\lambda d} (e^{-\lambda d} \lambda d) \right \}^N \label{eq:fork}
\end{equation}

In the proposed method, $N$ {\changed miniblocks} are required to be added
before adding a new block. Therefore, the above event has to succeed
$N$ times in order for a fork to {\changed occur}.

We now assume that all {\changed miniblocks} have the same creation rate
$\lambda = N$, where $N$ is the number of {\changed miniblocks} required to add
a block. Equation \ref{eq:fork} is plotted in
Fig.~\ref{fig:forkgraph}. It is shown that the probability of a fork can
be significantly lowered by increasing the number of {\changed miniblocks}.

\subsection{Variance of block time}

\begin{figure}[b]
\centering \includegraphics[width=0.8\columnwidth]
           {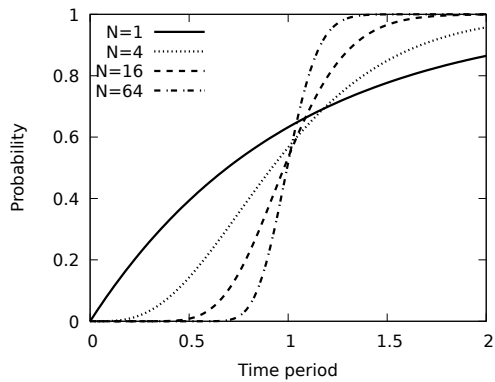}
\caption{Probability {\changed that} a block {\changed is} added within a time period with
  $N$ {\changed miniblocks}}
\label{fig:blocktime}
\end{figure}

Now we discuss how the variance {\changed in the} block time is
affected by introducing {\changed miniblocks}. The probability $b(t)$
{\changed that a block is} added within time period $t$ is the
probability {\changed that} $N$ or more {\changed miniblocks} are
added within this period. We have $N$ {\changed miniblock creations
  per interval on average; thus,}

\begin{equation}
b(t) = 1 - e^{-Nt}\sum_{i=0}^{N-1} \frac{(Nt)^i}{i!}.
\end{equation}

Fig. \ref{fig:blocktime} shows the probability {\changed that a block
  will} be added within a time period. The expected block time is 1,
and the variance is $1/N$. This means that the variance {\changed in
  the} block time decreases as the number of {\changed miniblocks}
increases.





\subsection{Winning probability of a node}

In Bitcoin, the winning probability for each miner is proportional to
its computational power. This is because when a block is added, the
probability {\changed that each node is} the node that added the block is
proportional to the number of hash values generated by that node.

Now, we discuss the winning probability of each node to add each
{\changed miniblock}. Suppose that there is no message delivery delay
and {\changed that} all nodes share the latest information. When a new
block is added, the total computation {\changed needed} for adding all
the corresponding {\changed miniblocks} is the same as the total
computational power spent by miners in that block time. The
probability $w_{n, k}$ {\changed that} node $n$ to {\changed adds
  miniblock} $k$ is as follows.

\[
w_{n, k} = \frac{\text{Computation power spent by $n$ for $k$}}{\text{Total
computational power spent for $k$}}
\]
\bigskip

\subsection{Properties of PoS}

The following is the list of properties that PoS has.

\begin{itemize}
\item All the properties of PoW listed in \ref{sec:bitcoin} are
  preserved.
\item Any node can submit a job and become a client.
\item A client can specify any instance of an
  optimization problem in a job.
\item A client can implement any search algorithm for any
  optimization problem for a job.
\item A client pays a charge for its job.
\item Miners have a financial incentive to find a good solution for
  the instance in a job.
\item Miners have a financial incentive to provide the best found
  solution to the client.
\item The charge is automatically paid to the node who provides the
  best solution to the client.
\item A probabilistic proof that the miners have evaluated a large
  number of solution candidates is provided.
\item Multiple jobs can be executed at a time.
\item The winning probability for each miner to find the best solution
  is proportional to the computational power spent by the miner for
  the job.
\item The expected amount of computation for a job is proportional to
  the charge paid for the job.
\item The variance {\changed in} block time is lower than that with
  PoW.
\item The probability of a fork is lower than that with PoW.
\item The storage capacity required to manage a blockchain based on
  PoS is not too large.
\item A blockchain based on PoS accepts jobs that run on computers
  with different architectures.
\end{itemize}

PoS has built-in countermeasures against the following cases.

\begin{itemize}
\item A node submits a problem instance {\changed for} which the node
  already knows a good solution.
\item A node has {\changed a} very effective way of searching for a solution
  {\changed to} some specific problems.
\item A node tries to steal a solution found by another node.
\end{itemize}

The requirements for an evaluator are as follows.

\begin{itemize}
\item It takes a solution candidate {\changed as} the first argument
  and a hash value {\changed as} the second argument.
\item It returns the evaluation of the solution candidate with a tiny
  amount of error depending on both the arguments.
\item It has to satisfy the condition explained in \ref{sec:loosely}.
\end{itemize}







\subsection{Security consideration}

Irrespective of search strategies, there is no difference in {\changed
  the} difficulty of generating a valid nonce. Thus, the following
items are allowed for nodes.

\begin{itemize}
\item Using its own search method, instead of the provided searcher.
\item Sharing intermediate results of a search among other nodes.
\item Participating in a search for only a particular job, instead of
  trying to add a whole block.
\item Starting the search {\changed immediately} after a job is
  submitted.
\end{itemize}
\bigskip

There is always {\changed a} possibility {\changed that} an evaluator
{\changed will} be reverse-engineered. If some malicious party comes
to know how error is introduced in some specific evaluator, they
{\changed could} try to quickly compute the output of {\changed the}
evaluator with a different second argument by reusing the result of
previous evaluation. This is a potential flaw {\changed in
  PoS. However, it} can be prevented by introducing {\changed the}
error in an early stage of evaluation rather than {\changed in} the
last step. For example, an evaluator for {\changed the} TSP could be
implemented to introduce error by slightly changing the positions of
cities.

When a {\changed miniblock} is added, the corresponding nonce with a
hash value beginning with {\changed the} required number of zero bits
is broadcast to the network. A solution candidate is included in this
nonce, and we can think of a cheat where a miner tries to find a
better solution by taking advantage of this solution candidate. There
is {\changed approximately} one block time to carry out a search after
a nonce is received before the deadline of registering a hash value
{\changed for} the found solution. However, such an attempt would be
hardly advantageous because this solution candidate is found in the
middle of a search, and therefore it is unlikely to be the best
solution found by the node that added the {\changed miniblock}. It is
also not very likely that the node that adds the block also finds the
best solution among all nodes.

{\changed
In another cheat, a client and miners can collude by submitting
problem instances which do not have a solution and forcing others to
work on these problems as the colluding miners work on instances
submitted by other clients. If a problem instance does not have a
solution, the corresponding evaluator always returns the same worst
evaluation value. If multiple miners find the best solutions with the
same evaluation value, the reward can be divided equally. This will
financially discourage this cheat because the colluding client will
have to pay other miners. }

We can think of another cheat where a malicious client implements an
evaluator in such a way that it returns {\changed an} unfairly high
evaluation when a special solution candidate is given. The client
would implement the searcher in such a way that it does not evaluate
such a solution candidate. By submitting a job with such an evaluator,
the client can almost always find the best solution and collect the
charge. However, this client cannot obtain the solutions found by
other nodes because a node publishes its solution only if it is the
winning node. The client is still be able to obtain the solution
candidate contained in the nonce. However, this would not be a very
good solution candidate, as explained above. Such a cheat can be
financially discouraged by giving a part of {\changed the} charge to
the node that adds the new block.

In order to load up the CPUs of rival miners, a client and a miner
might collude and submit jobs for which they already know a good
solution. In this case, there is no chance for rival miners {\changed
  to find} the best solution for the job since the colluded miner
will submit the best solution and collect the charge. This cheat can
be financially discouraged by giving a part of the charge to the node
that adds the new block. Rival miners {\changed will} have no
{\changed problems} in adding a new block with this cheat.

There are several ways of making a denial-of-service~(DoS) attack that
prevents execution or payment for {\changed jobs; although,} most of
them are not very effective. We can think of a DoS attack where a
malicious node registers a false good evaluation with a false hash
value on the blockchain when the best found solution is chosen and
provided to the client. This prevents the client from receiving the
true solution. If there is a node that makes such an attack, that node
has to be banned from the network and the payment process has to be
{\changed restarted.}

{\changed
\subsection{Applicable optimization problems}

Our protocol allows a blockchain to be used as a batch processing
system for solving optimization problems. For the society to have
demand for such a system, a large number of problem instances
satisfying the following two properties have to be provided. First,
the problem, the instance and the solution should not contain
privacy-sensitive data so that it is legal to publish them. Second,
solving an instance should require considerable amount of
computational resources. If a user can easily carry out the
computation on a small computer, there will be no reason to bother
executing the search on a batch processing system. The success of
BOINC projects~\cite{Anderson:2004:BSP:1032646.1033223} has
demonstrated that there is continued demand for such computation. The
computation involved in part of the BOINC projects and other
distributed computation projects shown below is optimization, and
therefore we believe that such projects can be ported to the computing
platform provided by our protocol.

Protein folding is simulated in Folding@Home~\cite{ZAGROVIC2002927} to
examine the causes of protein misfolding. To predict the folded
structure of a protein, a target protein sequence is deconstructed
into small fragments. Then, the qualities of fragments and their
assemblies are assessed by using some form of scoring function that
aims to select more native-like protein structures from among the many
possible combinations~\cite{Dill1042}.

MilkyWay@home~\cite{Newberg_2013} is a project that aims to generate
accurate three-dimensional dynamic models of stellar streams in the
immediate vicinity of the Milky Way. It runs two kinds of
applications. The first kind of application fits the spatial density
profile of tidal streams using statistical photometric parallax. The
other application finds the N-body simulation parameters that produce
tidal streams that best match the measured density profile of known
tidal streams.

In addition, many optimization problems have to be solved repeatedly
in fintech. For example, the Markowitz model is used for constructing
portfolios to optimize or maximize expected return based on a given
level of market risk~\cite{RePEc:bla:jfinan:v:7:y:1952:i:1:p:77-91,
  doi:10.1111/j.1540-6261.1970.tb00865.x}. Genetic algorithms are used
for index fund management~\cite{OH2005371}.

Optimization is also a key technique in deep
learning~\cite{Goodfellow-et-al-2016}, and we believe that there will
be continued demand for optimizing neural networks.  }

\section{Conclusion}

\label{sec:conclusion}

We have proposed a consensus protocol for blockchains with which the
wasted energy in the {\changed proof-of-work} system {\changed can} be used for
solving optimization problems submitted by any {\changed
  user}. Cryptocurrencies based on {\changed proof-of-work} are already very
popular, and to replace those cryptocurrencies, we designed our
protocol to be robust, secure and decentralized. Our protocol is
better than Gridcoin in that sense since our protocol does not depend
on any external organization or system.

{\changed
The motivation of this research is to apply the huge amount of
electricity and computational power to useful work. The proposed
system automatically adjusts the charge according to the balance
between the demand and the total amount of computing power provided by
the miners. If the demand is very small, the price for the
optimization service will drop significantly. In such a situation, we
can expect an increase in demand according to the law of
supply. Therefore, fallback to PoW should not happen frequently, and
the proposed method is expected to effectively prevent energy from
being wasted.
}

There will be three kinds of users of a PoS-based blockchain, and
these three user groups can be independent of each other. The first
kind of users {\changed is} those who want to use a PoS-based
blockchain as a payment method. The second kind of users {\changed is}
those who spend their computational resources to earn e-coins. The
existing blockchains already have these two kinds of users, and PoS
adds {\changed a} third kind of {\changed user who uses} a PoS-based
blockchain as a grid computing infrastructure. The computational
service provided by our protocol would be beneficial for computation
tasks whose intermediate results can be published. Considering
{\changed the popularity of} BOINC projects and public clouds, we
expect that there is a large public demand for such a computational
service.

The computation in our protocol is ASIC resistant. This would make
mining with ordinary CPUs more profitable. Unlike cloud computing,
{\changed the} computers used in mining need not be reliable. Since
the computational resources are {\changed no longer wasted,} public
organizations and more general users could be expected to join. This
would make a blockchain with our protocol more decentralized than
existing ones.



\section*{Acknowledgment}

The authors would like to thank Prof. Mike Barker for his
suggestions. {\changed We would like to thank AJE for editing and reviewing this
manuscript for English language.}

\ifCLASSOPTIONcaptionsoff
  \newpage
\fi



%
\bibliography{posearch}
\bibliographystyle{IEEEtran}

%

\begin{IEEEbiography}{Naoki Shibata}
is an associate professor at Nara Institute of Science and
Technology. He received the Ph.D. degree in Computer Science from
Osaka University, Japan, in 2001. He was an assistant professor at
Nara Institute of Science and Technology 2001-2003 and an associate
professor at Shiga University 2004-2012. His research areas include
distributed systems, inter-vehicle communication, mobile computing,
multimedia communication, and parallel algorithms. He is a member of
IPSJ, ACM and IEEE.
\end{IEEEbiography}

\EOD

\end{document}